\documentclass[a4paper,12pt]{article}
\begin{document}
\vbadness = 100000
\hbadness = 100000

\title{Comment on Jerrold Franklin, Superluminal neutrinos, arXiv:1110.0234v1}
\author{Jacques Goldberg\footnote{Internet address:
Jacques.Goldberg@cern.ch}\\
Department of Physics\\
Technion-Israel Institute of Technology, Haifa, Israel
\date{\today}}
\maketitle
\begin{abstract}A reasonably short antineutrino run using OPERA and/or MINOS may
test J.~Franklin's explanation of recently claimed $\nu_\mu$ superluminality.
\end{abstract}

J.~Franklin has just shown\cite{franklin} that when a neutrino beam 
travels in the crust of the Earth, as in the MINOS\cite{minos} and OPERA\cite{opera} experiments, if the imaginary part of a Lorentz scalar potential related to the attenuation is larger than the mass of the neutrino, this leads naturally to superluminal propagation.

It is well known\cite{kim-pevsner} that neutrinos and antineutrinos behave quite
differently in matter. Whatever the Lorentz scalar potential behind 
Franklin's conjecture, a relatively short exposure of the MINOS or OPERA,
 or even better, both, detectors to a $\bar{\nu}_{\mu}$ beam may support the conjecture, if the superluminality is found to differ for neutrinos and antineutrinos (presumably reduced for neutrinos).

Switching the beam polarity is definitely very far from trivial, but may 
efficiently help discard a number of alternative explanations for a surprising observation.


\begin{thebibliography}{9}
\bibitem{franklin}J.~Franklin, Superluminal neutrinos, arXiv:1110.234v1
\bibitem{minos}P.~Adamson {\em et al.}, Phys.Rev.D76 (2007) 072005.
\bibitem{opera}Opera Collaboration, {\it Measurement of the neutrino velocity with the OPERA detector in the
CNGS beam, arXiv:1109.4897}.
\bibitem{kim-pevsner}C.W.~Kim and A.~Pevsner, Neutrinos in Physics and Astrophysics,Chapter 8, Harwood Academic Publishers, 1993, ISBN 3-7186-0567-8. 
\end{thebibliography}
\end{document}